\global\def\draftcontrol{0}

\def\versionno{ time -- draft -- 11.13.03  }
\catcode`\@=11

\expandafter\ifx\csname draftcontrol\endcsname\relax\global\def\draftcontrol{0}
\fi

{\count255=\time\divide\count255 by 60
\xdef\hourmin{\number\count255}
\multiply\count255 by-60\advance\count255 by\time
\xdef\hourmin{\hourmin:\ifnum\count255<10 0\fi\the\count255}}
\def\draftdate{\number\month/\number\day/\number\year\ \ \ \hourmin }

\newcommand\makepapertitle{\par
  \begingroup
    \renewcommand\thefootnote{\@fnsymbol\c@footnote}%
    \def\@makefnmark{\rlap{\@textsuperscript{\normalfont\@thefnmark}}}%
    \long\def\@makefntext##1{\parindent 1em\noindent
            \hb@xt@1.8em{%
                \hss\@textsuperscript{\normalfont\@thefnmark}}##1}%
     \newpage
     \global\@topnum\z@   % Prevents figures from going at top of page.
     \@makepapertitle
     \thispagestyle{empty}\@thanks
  \endgroup
  \setcounter{footnote}{0}%
  \global\let\thanks\relax
  \global\let\makepapertitle\relax
  \global\let\@makepapertitle\relax
  \global\let\@thanks\@empty
  \global\let\@author\@empty
  \global\let\@date\@empty
  \global\let\@title\@empty
  \global\let\title\relax
  \global\let\author\relax
  \global\let\date\relax
  \global\let\and\relax
  \def\version{\let\version\@version\@gobble}
}
\def\@makepapertitle{%
  \newpage
   \ifnum\draftcontrol=1 {}
   \version\versionno
   \vskip 3em%
   \else
   \hfill\hbox to 3cm {\parbox{4cm}{\@pubnum}\hss}%
   \vskip 3em%
   \fi
   \begin{center}%
   \let \footnote \thanks
     {\LARGE \@title \par}%
     \vskip 1.5em%
     {\normalsize%\large
       \lineskip .5em%
       \begin{tabular}[t]{c}%
         \@author
       \end{tabular}\par}%
     \vskip 1em%
     {\@bstract}%
     \end{center}%
     \vskip .5em
     \@date%
   \par
}

\gdef\@pubnum{}
\def\pubnum#1{%
  \gdef\@pubnum{#1}}

\gdef\@bstract{}
\def\Abstract#1{%
  \gdef\@bstract{%
   \parbox{\textwidth-0pc}{%
   \centerline{\bf Abstract}\penalty1000
   \noindent%\abstractfont \baselineskip=12pt
   \renewcommand\baselinestretch{1.0}
   {#1}}}
}

\def\ps@paper{\let\@mkboth\@gobbletwo%
     \ifnum\draftcontrol=1
        \def\@oddfoot{\hbox to \textwidth{\tiny \versionno \hfil\tiny\draftdate}%
        \hskip -\textwidth \hbox to \textwidth{\hfil\rm\thepage\hfil}}%
     \else\def\@oddfoot{\hbox to \textwidth{\hfil\rm\thepage\hfil}}
     \fi
     \let\@evenfoot\@oddfoot
}
\def\body{\clearpage
          \pagestyle{paper}
        }
\newenvironment{acknowledgments}{%
\vskip 3.25ex
\noindent {\bf Acknowledgments}
}

\def\@version#1{\ifnum\draftcontrol=1
\typeout{}\typeout{#1}\typeout{}
\vskip3mm\centerline{\hbox{\fbox{\normalsize{\tt DRAFT -- #1 -- }
                   {\draftdate}}}}\vskip3mm
\fi}
\let\version\@version
\long\def\eqlabel#1{\ifnum\draftcontrol=1
                    \tag@false  % there are some problems with multline without this
                    \tag*{(\theequation) \hbox to -0.2cm{\hspace{0cm}\small{#1}\hss}}
                    \refstepcounter{equation} 
                    \edef\@currentlabel{\theequation}
                    \ltx@label{#1}          % use old LaTeX \label instead of new definition
                    \else
                    \label{#1}
                    \fi
                    }

\documentclass[12pt,letterpaper]{article}
\usepackage{amsmath,amssymb,array,calc,rotating,epsfig,psfrag}
\usepackage[nosort]{cite}
\ifnum\draftcontrol=1
\fi
\renewcommand\baselinestretch{1.25}
\renewcommand\section{\@startsection {section}{1}{\z@}%
                                   {-3.5ex \@plus -1ex \@minus -.2ex}%
                                   {2.3ex \@plus.2ex}%
                                   {\normalfont\large\bfseries}}
\renewcommand\subsection{\@startsection{subsection}{2}{\z@}%
                                     {-3.25ex\@plus -1ex \@minus -.2ex}%
                                     {1.5ex \@plus .2ex}%
                                     {\normalfont\normalsize\bfseries}}
\renewcommand\subsubsection{\@startsection{subsubsection}{3}{\z@}%
                                     {-3.25ex\@plus -1ex \@minus -.2ex}%
                                     {1.5ex \@plus .2ex}%
                                     {\normalfont\normalsize\it}}

\numberwithin{equation}{section}

\def\projective   {{\mathbb P}}

\def\zet          {{\mathbb Z}}

\def\del          {\partial}

\def\be		{\begin{equation}}
\def\ende		{\end{equation}}

\def\revise#1       {\marginpar{\rule{2mm}{1cm} #1}}

\def\ZZ{\zet}

\newcommand{\nc}{\newcommand}

\def\bea		{\begin{eqnarray}}
\def\eea		{\end{eqnarray}}
\nc{\e}{{\rm exp}}

\nc{\cosech}{{\rm cosech}}

\nc{\Li}{{\rm Li_{2}}}

\nc{\li}{\lambda_{i}}
\nc{\lj}{\lambda_{j}}
\nc{\lk}{\lambda_{k}}
\nc{\laml}{\lambda_{l}}

\nc{\mi}{\mu_{i}}
\nc{\mj}{\mu_{j}}
\nc{\mk}{\mu_{k}}
\nc{\ml}{\mu_{l}}

\nc{\om}{\omega}

\def\ep{\epsilon_2}
\def\epp{\epsilon_1}
\def\A{{\cal A}}
\def\B{{\cal B}}

\nc{\non}{\nonumber}

\catcode`\@=12

\begin{document}

\title{The Spectral Curve of the Lens Space Matrix Model} 

\pubnum{%
USC-03-09 \\
NSF-KITP-03-109 \\
hep-th/0311117}

\date{November 2003}

\author{Nick Halmagyi\footnote{halmagyi@physics.usc.edu}$\ ^{1,2}$
and Vadim Yasnov\footnote{yasnov@physics.usc.edu}$\ ^{1}$\\[0.4cm]
\it $^{1}$Department of Physics and Astronomy\\
\it University of Southern California \\
\it Los Angeles, CA 90089, USA \\[0.2cm]
\it $^{2}$Kavli Institute for Theoretical Physics\\
\it University of California\\
\it Santa Barbara, CA 93106, USA
}

\Abstract{
Following hep-th/0211098 we study the matrix model which describes the topological A-model on
$T^{*}(S^{3}/\ZZ_p)$. We show that the resolvent has square root branch cuts and it follows that this is a $p$ cut single matrix model. We solve for the resolvent and find the spectral curve. We comment on how this is related to large $N$ transitions and mirror symmetry.
}

\enlargethispage{1.5cm}

\makepapertitle

\body

%------------------------------------------------------------------------------------------------------------------------

\section{Introduction}

The duality between open and closed string theories is a fascinating area of string theory. This duality is often understood as a geometric transition, where topologically distinct manifolds are used for the open or closed string theory, the prototypical example being the duality between the resolved and deformed conifolds  \cite{Gopakumar:1998ki, Klebanov:2000hb, Cachazo:2001jy} . At the level of topological string theory, for the A-model transition the closed string side is the resolved conifold and the open string side is the deformed conifold, for the B-model transition this is reversed. 

The study of the B-model conifold transition and its generalizations led to the introduction of matrix models as a way to describe holomorphic Chern-Simon's (HCS) theory reduced to the $2$-cycles of the generalized conifold \cite{Dijkgraaf:2002fc}. This is in turn directly related to four dimensional ${\cal N}=1$ Yang-Mills theory, because the partition function of the topological string which was used to engineer the Yang-Mills theory gives the low energy effective superpotential \cite{Bershadsky:1993cx, Dijkgraaf:2002dh}. This is now known as Dijkgraaf-Vafa (DV) theory. The connection between matrix models and superpotentials has also been uncovered directly in the field theory \cite{Cachazo:2002ry, Dijkgraaf:2002xd}.

Matrix models were introduced into the topological A-model from a very different standpoint by Marino \cite{Marino:2002fk}, where he presented a matrix model description of Chern-Simons (CS) theory on certain 3-manifolds. This matrix model always has a quadratic potential, but it has a rather strange measure which encodes the different geometries, when this 3-manifold is $S^{3}$ this is the Haar measure on $SU(N)$. This work was extended in \cite{Aganagic:2002wv} where they considered the A-model open topological string on $T^{*}(S^{3}/\ZZ_{p})$ (corresponding to CS theory on $S^{3}/\ZZ_{p}$ \cite{Witten:1992fb}) and also the mirror geometry ($\widetilde{X}$). By using similar reasoning as in \cite{Dijkgraaf:2002fc} they were able to derive a matrix model for HCS theory reduced to $\projective^{1}$'s in $\widetilde{X}$. As expected but still quite remarkably, for each $p$ HCS on $\widetilde{X}$ and CS theory on $S^{3}/\ZZ_{p}$ are described by identical matrix models. Many of the ideas at work here (pre matrix model) are covered in the great review paper by Marino \cite{Marino:2002wa}.

So essentially, by studying the topological A-model and using mirror symmetry, Dijkgraaf-Vafa (DV) theory was extended to a new class of Calabi-Yau manifolds. Now by the general principles of DV theory, special geometry on the closed string dual geometry of $\widetilde{X}$ (call it $X$), should reduce to special geometry on a Riemann surface in $X$, and this surface should be the spectral curve of the aforementioned matrix model. For the case of the A-model on $T^{*}S^{3}$, the spectral curve was shown to coincide with the non-trivial Riemann surface in $X$ and the leading order (in $g_{s}$) free energy of the matrix model was shown to agree with the known result. In \cite{Tierz:2002jj} the free energy of this matrix model was calculated to all orders and shown to agree with known results \cite{Gopakumar:1998ki}. In \cite{Halmagyi:2003fy} the orientifold of the conifold was considered and the subleading order free energy was shown to agree with known results \cite{Sinha:2000ap}.
 
In this paper we investigate the matrix model of CS theory on $S^{3}/\ZZ_{p}$. It was shown in \cite{Aganagic:2002wv} that this matrix model has $p$ cuts each at the position of a $\projective^{1}$ in the blown up 3-fold and it was noticed that this model looks similar to a $p$-matrix model. We will show that the resolvent has square root cuts which implies that really it is a single matrix model with $p$ cuts. We then find that the spectral curve is a genus ($p-1$) Riemann surface with four points deleted and find the equation for this curve. For the case of $p=2$ we compare our surface to that obtained from the Hori-Vafa mirror. 

This paper is organized as follows. In section 2. we discuss the geometrical structures which are involved in the large $N$ duality we are considering and in mirror symmetry of these dualities. In section 3 we review the solution of the matrix model for CS theory on $S^{3}$ and also solve it with our new method. In  sections 4 and 5 we solve the case of $S^{3}/\ZZ_{2}$ and $S^{3}/\ZZ_{p}$ respectively. In section 6. we outline our calculation of the free energy for $S^{3}/\ZZ_{2}$ which we view as a non-trivial check of our method, the full calculation is presented in the appendix.

%--------------------------------------------------------------------------------------------------------------------

\section{Geometry}

The first geometric transition to be studied was the A-model conifold transition of Gopakumar and Vafa \cite{Gopakumar:1998ki}. They considered the closed topological A-model on the resolved conifold (${\cal O}_{-1}\hspace{0.2mm} +\hspace{0.2mm}{\cal O}_{-1}\! \rightarrow\! \! \projective^{1}$) and argued that it is equivalent to the open topological A-model on the deformed conifold ($T^{*}(S^{3})$). This has been extended significantly to a large class of toric Calabi-Yau's \cite{Aganagic:2002qg, Diaconescu:2002sf, Diaconescu:2002qf} but not including the expected transition between $A_{p-1}\rightarrow \projective^{1}$ and $S^{3}/\ZZ_{p}$.

Heuristically, taking a $\ZZ_{p}$ orbifold on both sides of the A-model conifold transition should produce a transition between some $A_{p-1}$ fibration over $\projective^{1}$ on the closed string side and $T^{*}(S^{3}/\ZZ_{p})$ on the open string side. In \cite{Aganagic:2002wv} the matrix model of CS on $S^{3}/\ZZ_{2}$ was studied, its free energy was calculated perturbatively and was shown to agree with the closed A-model on ${\cal O}(-K)\! \! \rightarrow\! \! \projective^{1}\times\projective^{1}$ (which is a trivial $A_{1}$ fibration over $\projective^{1}$), thus implying that $T^{*}(S^{3}/\ZZ_{2})$ undergoes a geometric transition to ${\cal O}(-K)\! \! \rightarrow\! \! \projective^{1}\times\projective^{1}$. For the case of $T^{*}(S^{3}/\ZZ_{p})$ the Riemann surface embedded in $X$ is given by the Hori-Vafa mirror map, which we will now briefly describe. 

By studying T-duality on $T^{3}$ fibres of an arbitrary toric Calabi-Yau manifold \cite{Hori:2000kt}, a mirror map was derived. This map can be reduced to the following operation. Take a toric web diagram\footnote{a toric web diagram is a trivalent graph such that three unit vectors emenating from each node sum to zero \cite{Leung:1997tw}, it encodes the singular structure of the $T^{3}$ fibration.} of a toric CY threefold $M$, then consider the Riemann surface obtained by thickening each line into a cylinder. This Riemann surface will be 

\be
F(e^{u},e^{v})=0,
\ende
and then the 3-fold mirror to $M$ is given by 

\be
xy=F(e^{u},e^{v}).
\ende
The Hori-Vafa mirror map gives $F(e^{u},e^{v})$ explicitly.
\begin{figure}[bth!]
\centerline{ \epsfig{file=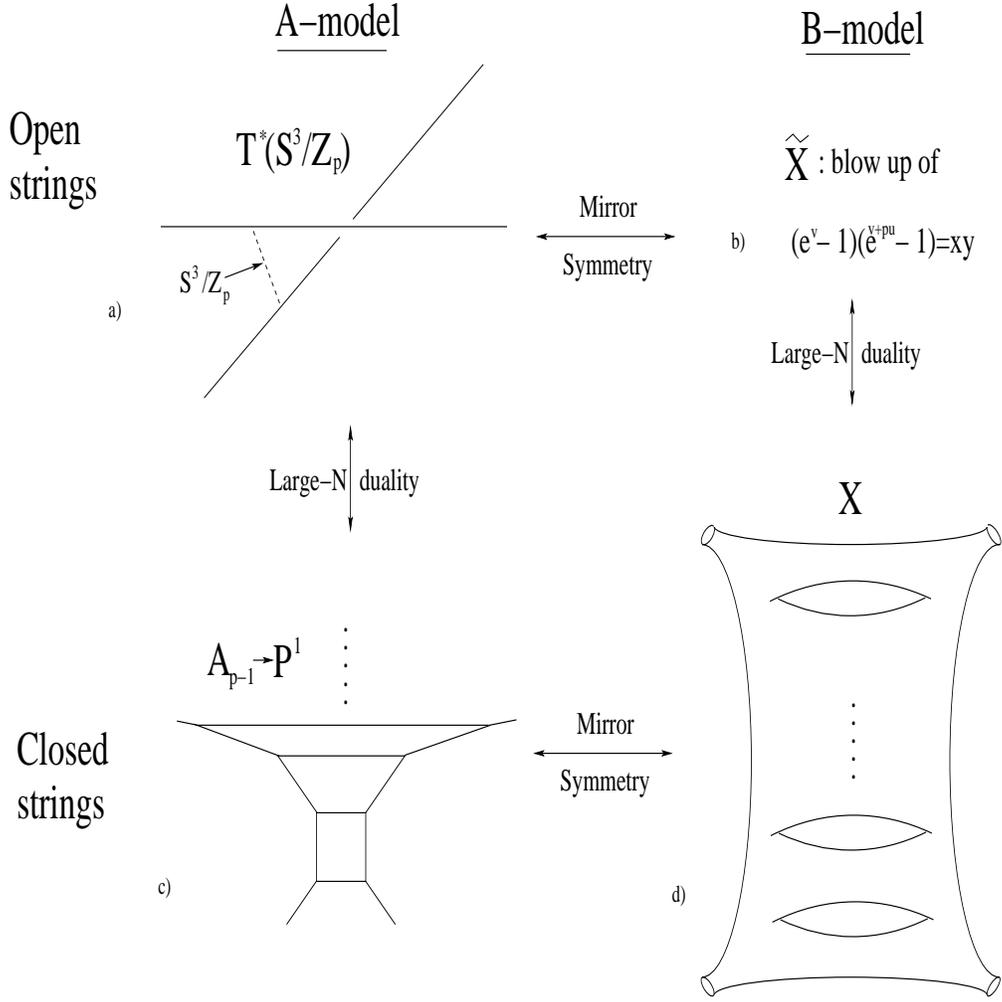,width=14cm,height=14cm}}
\caption{\sl Large N dualities and mirror symmetry. 
a) $T^{*}(S^{3}/\ZZ_{p})$ represented by a deformation of a toric diagram \cite{Aganagic:2001ug}  b) The mirror to $T^{*}(S^{3}/\ZZ_{p})$  c) Schematic picture of the  toric web of an $A_{p-1}$ fibration over $\projective^{1}$  d) The Hori-Vafa mirror map gives a bundle over a genus $p-1$ Riemann surface, where the Riemann surface is simply a thickening of the toric web diagram   }
\label{riemann}
\end{figure}
The various geometries and dualities involved here are shown in Fig \ref{riemann}.

This map gives the mirror to the resolved conifold (${\cal O}_{-1}\hspace{0.2mm} +\hspace{0.2mm}{\cal O}_{-1}\! \rightarrow\! \! \projective^{1}$) to be

\be \label{conifold}
xy=(e^{v}-1)(e^{u+v}-1)-1+e^{t}\equiv F_{c}(e^{u},e^{v})
\ende
and the spectral curve of the appropriate matrix model (CS theory on $S^{3}$) was shown in \cite{Aganagic:2002wv} to be given by $F_{c}(e^{u},e^{v})=0$. In this paper we further study this by reconsidering the general case of CS theory on $T^{*}(S^{3}/\ZZ_{p})$. The mirror ($\widetilde{X}$) of $T^{*}(S^{3}/\ZZ_{p})$ is given by blowing up the singular 3-fold

\be \label{zpCY}
xy=(e^{v}-1)(e^{v+pu}-1)\equiv F_{p}(e^{u},e^{v}).
\ende
We will find that the associated matrix model has a spectral curve which is a genus ($p-1$) Riemann surface with four points deleted, given by a certain complex structure deformation of $F_{p}(e^{u},e^{v})=0$. Whilst we cannot calculate precisely the complex structure parameters, for the case $p=2$ we can give an expansion in the 't Hooft parameters, which in principle could be generalized to $p>2$. We also find which monomials appear in the deformation. All this relies on our showing that the matrix model is a single matrix model with $p$ square root cuts\footnote{In \cite{Aganagic:2002wv} it was suggested that one could view it as a $p$-matrix model, this would produce a $p$-sheeted Riemann surface as for the quiver matrix model \cite{Dijkgraaf:2002vw, Hofman:2002bi, Lazaroiu:2003vh}.}.

%--------------------------------------------------------------------------------------------------------------------
\section{Chern-Simons matrix model on $S^3$.}

In this section we review the  matrix model that describes CS theory on $S^3$ \cite{Aganagic:2002wv}. We study this model with a new method we have developed,
a method which generalizes nicely to $S^3/\ZZ_p$

%---------------------------------------------------------------------------------------------------------------------------------

\subsection{Solution by contour integral}
The matrix integral is given by 

\be
\label{eqmS3}
Z \sim \int\prod_i du_i\Delta^2 (u) \exp \left (-\frac{1}{g_s} \sum_i u^2_i/2\right )
\ende 
where the group measure is an analytic continuation of the Haar measure,

\be
\label{measure}
\Delta (u)=\prod_{i<j}2\sinh \left (\frac{u_i-u_j}{2}\right ).
\ende 
Although the measure is periodic, the potential is not, therefore the domain of
integration is non-compact.
The equation of motion for each eigenvalue is

\be 
\label{EM}
\frac{1}{g_s} u_i=\sum_{j\neq i}\coth \left (\frac{u_i-u_j}{2}\right ).
\ende 
 
In general, the form of the resolvent can be inferred from the function on the r.h.s. of the equations of motion and the measure (\ref{measure}). In this case it is

\be
\label{omegadef}
\omega (z)=g_s\sum_i\coth \left (\frac{z-u_i}{2}\right ).
\ende 
We then multiply (\ref{EM}) by $\coth ((z-u_i)/2)$, sum over eigenvalues  and take the large N
limit. This leads to the following loop equation

\be
\label{EMlargeN}
{\left(\frac{\omega (z)}{2}\right)}^2-z\frac{\omega (z)}{2}=f(z)+\frac{1}{4} S^2,
\ende 
with 

\be
f(z)=\frac{1}{2}g_s\sum_i (u_i-z)\coth\left (\frac{z-u_i}{2}\right )
\ende 
being a regular function. Eq (\ref{EMlargeN}) shows that the resolvent acquires a square root
cut in the large N limit and that there is only one cut.

The spectral curve is obtained by gluing two infinite cylinders along the cut. There are two independent cycles. The $A$ cycle is a contour around the cut, the $B$ cycle starts at infinity on the classical sheet where the resolvent is finite and goes to the other sheet through the cut. We call $S=g_s N$ the 't Hooft parameter. From equation (\ref{omegadef}), we get the limiting value of the resolvent,

\be
\lim_{z\rightarrow \infty}\omega(z)=S
\ende
One also has to fix the period over the A-cycle

\be
\pi i S=\oint_A\frac{\omega (z)}{4}dz.
\ende 
The last condition is equivalent to $\widetilde{\omega}(z)$, the other branch of the resolvent, having the limiting value

\be
\lim_{z\rightarrow -\infty}\widetilde{\omega}(z)=-S.
\ende
This agrees with the definition (\ref{omegadef}). All these conditions are sufficient to find the resolvent. We review briefly how it is done in \cite{Aganagic:2002wv}.
 
The equation of motion can be written a little differently by introducing a new resolvent

\be
v(Z)\equiv g_s\sum_i\frac{U_i}{U_i-Z}
\ende 
with $U_i=e^{u_i}$ and $Z=e^z$. Importantly, both $v(Z)$ and $\omega(Z)$ have the same singular behaviour, the relation between them is given by 

\be
\omega (Z)=S-2v(Z).
\ende 
The problem is now essentially a Hermitian matrix model with a logarithmic potential, which leads by standard arguments \cite{Kazakov:1996} to

\be
\label{contour}
-2 v(Z)=\sqrt{(Z-a) (Z-b)}\oint_C\frac{dX}{2\pi i}\frac{\log (X e^{-S})}{X-Z}\frac{1}{\sqrt{(X-a) (X-b)}},
\ende 
where the contour $C$ encircles the cut but not the point $Z$. The normalization conditions 
at $\pm\infty$ fix the end points of the cut and the final answer is

\be
\label{vafa'somega}
\omega (Z)=\log \left (\frac{e^{-S/2}}{2}\left (Z+1-\sqrt{(1+Z)^2-4 Z e^S}\right )\right ).
\ende 
The spectral curve is the surface where the resolvent is well defined, in this case it is given by (as advertised after eq. \ref{conifold})

\be \label{s3spec}
(e^{v}-1)(e^{u+v}-1)+e^{S}-1=0,
\ende
where, $u\equiv z$. 

Although this procedure is not very lengthy it becomes difficult even for the case of $S^3/\zet_2$ 
lens space. 

%---------------------------------------------------------------------------------------------------------------------------------

\subsection{Solution by a regular function}

We have found another way of finding the resolvent, this method will easily generalize to the case of $S^3/\ZZ_p$. 

Let $\omega_+$ be the value of the resolvent on one edge of the cut and $\omega_-$ be the value of the resolvent on the other edge. From the large $N$ limit of (\ref{EM}), it is clear that 

\be
\frac{\omega_+ (z)}{2}+\frac{\omega_- (z)}{2}=z.
\ende 
We then construct the function 

\be \label{gdef}
g(Z)\equiv e^{\omega/2}+Ze^{-\omega/2}.
\ende  
which is regular everywhere except at infinity. The limiting behavior of the resolvent will completely determine this function, 

\bea
&& \lim_{Z\rightarrow \infty} g(Z) =e^{-S/2}Z, \\
&& \lim_{Z\rightarrow 0} g(Z) =e^{-S/2}.
\eea
The unique function that satisfies these conditions is 

\be
g(Z)=e^{-S/2} (Z+1).
\ende 

Now the quadratic equation (\ref{gdef}) gives the resolvent
explicitly,

\be
e^{\omega/2}=\frac{1}{2}\left (g(Z)-\sqrt{g^2(Z)-4 Z}\right ).
\ende 
This is the same resolvent as (\ref{vafa'somega}) that is obtained using the contour integral representation and thus we get the same spectral curve (\ref{s3spec}). 

Let us summarize the main strategy. From the large $N$ limit of the equation of motion we can deduce that the resolvent has a square root cut, then the value of the resolvent on one edge of the cut can be simply related to the value of the resolvent on the other edge of the cut. Knowing this, one has to construct a function of the resolvent that is regular everywhere except infinity. The main ingredients are the functions $e^{\omega/2}$ and $e^{-\omega/2}$. Once such a function is found, $e^{\omega/2}$ can be written as a solution to a quadratic equation. This strategy will be shown to work for all Lens spaces. For $S^{3}/\ZZ_p$ with $p$ even, it is also possible to construct a function which is square root branched on each cut and from this, solve for $\omega(Z)$.

%---------------------------------------------------------------------------------------------------------------------------------
\section{$S^3/\ZZ_2$ Lens space resolvent.}

We now employ the strategy from the previous section for the geometry $S^3/\ZZ_2$. We refer the reader to \cite{Aganagic:2002wv} for a derivation of the Lens space matrix model but the reader can also just take (\ref{z2partition}) as a starting point. The partition function for CS theory on $S^3/\ZZ_2$ is given by the integral over two sets of eigenvalues 

\be \label{z2partition}
Z \sim \int\prod_i du_i\prod_\alpha d\mu_\alpha\Delta^2 (u,\mu)\exp \left(-\frac{1}{g_s} V(u,\mu)\right ),
\ende
where the measure is

\be
\Delta (u,\mu)=\prod_{i<j}2\sinh\left (\frac{u_i-u_j}{2}\right )
\prod_{\alpha < \beta}2\sinh\left (\frac{\mu_\alpha-\mu_\beta}{2}\right )\prod_{i,\alpha}2\cosh\left (\frac{u_i-\mu_\alpha}{2}\right )
\ende
and $i\! \in\! (1,N_1)$, $\alpha\! \in\! (1,N_2)$. Anticipating taking the large $N$ limit we also 
introduce two 't Hooft parameters $S_1=g_s N_1$ and $S_2=g_s N_2$ and $S=S_1+S_2$. The potential is 

\be
V(u,\mu)=\left( 2\sum_i u_i^2+2\sum_\alpha\mu_\alpha^2\right )/2,
\ende
and the equations of motion for each eigenvalue are

\bea \label{p2eom1}
&&  2 u_{i}= g_{s}\sum_{j\neq i}\coth \left(\frac{u_{i}-u_{j}}{2}\right) + 
g_{s}\sum_{\alpha}\tanh \left(\frac{u_{i}-\mu_{\alpha}}{2}\right) \\
&& 2 \mu_{\alpha}= g_{s}\sum_{\beta \neq \alpha}\coth \left(\frac{\mu_{\alpha} - \mu_{\beta}}{2}\right) + 
g_{s}\sum_{i}\tanh \left(\frac{\mu_{\alpha}-u_{i}}{2}\right) \label{p2eom2}.
\eea
We define the resolvents as

\be \label{defw}
\omega (z)=g_s\sum_i\coth\left (\frac{z-u_i}{2}\right )+g_s\sum_\alpha\tanh\left (\frac{z-\mu_\alpha}{2}\right ).
\ende
\be
\omega_1 (z)=g_s\sum_i\coth\left (\frac{z-u_i}{2}\right ),
\ende
\be
\omega_2 (z)=g_s\sum_\alpha\coth \left (\frac{z-\mu_\alpha}{2}\right ),
\ende
so the relation between them reads
\be
\label{ww1w2}
\omega (z)=\omega_1 (z)+\omega_2 (z-i\pi).
\ende

%---------------------------------------------------------------------------------------------------------------------------------

\subsection{Solution by a regular function}

Now we multiply equation (\ref{p2eom1}) by `$\coth ((z-u_i)/2)$' and sum over $i$, as well as multiplying equation (\ref{p2eom2}) by `$\tanh ((z-\mu_\alpha)/2)$' and summing over $\alpha$. Then we add these two equations 
and take the large $N$ limit, with the result being

\be \label{EMN}
{\left (\frac{\omega (z)}{2}\right )}^2-2z\frac{\omega_1 (z)}{2}-2(z-i\pi)\frac{\omega_2 (z-i\pi)}{2}=f(z)
\ende
where

\be
f(z)=g_s\sum_i (u_i-z)\coth \left (\frac{z-u_i}{2}\right )+g_s\sum_\alpha (\mu_\alpha-(z-i\pi))\tanh\left (\frac{z-\mu_\alpha}{2} \right )+\frac{1}{4}S^2
\ende
is a regular function. We can write (\ref{EMN}) in two ways,

\bea
&& {\left (\frac{\omega (z)}{4}\right )}^2-(z-i\pi)\frac{\omega (z)}{4}-i\pi\frac{\omega_1 (z)}{4}=\frac{f(z)}{4}, \\
&& {\left (\frac{\omega (z+i\pi)}{4}\right )}^2-(z+i\pi)\frac{\omega (z+i\pi)}{4}+i\pi\frac{\omega_2 (z)}{4}=\frac{f(z+i\pi)}{4}.
\eea
Now we make an important assumption, we assume that the eigenvalues spread only along the real line. For general multi matrix models this is not true \cite{Dijkgraaf:2002vw, Hofman:2002bi, Lazaroiu:2003vh}. However as we will see this assumption leads to the correct result for our case. It follows that if $\omega_1 (z)$ jumps at a point $z$ then $\omega_2 (z-i\pi)$ does not and vice versa. Note that we do not make any assumption on the type of the cuts. In the total resolvent $\omega (z)$, the individual resolvents come with a relative shift of the argument by $i\pi$. Therefore the two cuts in the total resolvent are now separated by $i\pi$. On one cut the total resolvent jumps only due to $\omega_1(z)$ and on the other cut only due to $\omega_2(z)$. From this we can deduce that 

\bea \label{p2eom3}
&& \frac{1}{4}\left( \omega_+(z)+\omega_{-}(z)\right) =z \ \ \ \ ({\rm u\ cut})\\
&& \frac{1}{4}\left( \omega_+(z+i\pi)+\omega_{-}(z+i\pi)\right) =z  \ \ \ \ ({\rm \mu\ cut})\label{p2eom4}
\eea
and so the resolvent $\omega (z)$ really does have square root branch cuts\footnote{due to the fact that the matrix model looks much like a 2-matrix model, the concern was that it may be branched by a cubic  root.}. Using (\ref{p2eom3}, \ref{p2eom4})\footnote{which should be thought of as a principle value integral.}, it is straightforward to find a function of $\omega(z)$ which is regular everywhere except at infinity, it is 

\be\label{reg}
g(Z)\equiv e^{\omega/2}+Z^2e^{-\omega/2}.
\ende
This function is regular and has limiting behavior

\bea
&& \lim_{Z\rightarrow \infty} g(Z)=Z^{2}e^{-S/2}, \\
&& \lim_{Z\rightarrow 0} g(Z)=e^{-S/2}.
\eea
Therefore $g(Z)$ can be written in terms of only one unknown parameter 

\be \label{gp2}
g(Z)=e^{-S/2}(Z^2+d Z+1),
\ende
where $d$ is related to the end points of the cuts. 

Solving (\ref{reg}) as a quadratic equation for $e^{\omega/2}$ yields,

\be \label{w/2}
\frac{\omega (Z)}{2}=\log\left (\frac{1}{2}\left (g(Z)-\sqrt{g^2(Z)-4 Z^2}\right )\right ).
\ende
It is easy to see that $\frac{1}{2}(\omega_{+}(Z) + \omega_{-} (Z))=\log (Z^2)$ and therefore (\ref{p2eom3})
and (\ref{p2eom4}) are satisfied. 

Now consider the function under the square root sign in (\ref{w/2}). If $Z_i$ sets this to zero, then $1/Z_{i}$ will as well. Together with fact that the eigenvalues are all real, this implies that the end points of each cut are the inverse of one another, i.e. the $u$ cut is $(a,1/a)$, the $\mu$ cut is $(b,1/b)$ for some $a,b$. Further, the relationship between our parameter $d$ and the end point of the cuts is easy to find

\bea 
&& d=2 e^{S/2}-\left (a+\frac{1}{a}\right ), \\
&& d=-2 e^{S/2}+\left (b+\frac{1}{b}\right ). \label{xb}
\eea

As discussed in the previous section, the spectral curve is two cylindrical sheets glued together along these cuts. The center of the $u$ cut is at $z=0$ and the center of the $\mu$ cut is located at $z=i\pi$. 

\begin{figure}[bt]
\centerline{ \epsfig{file=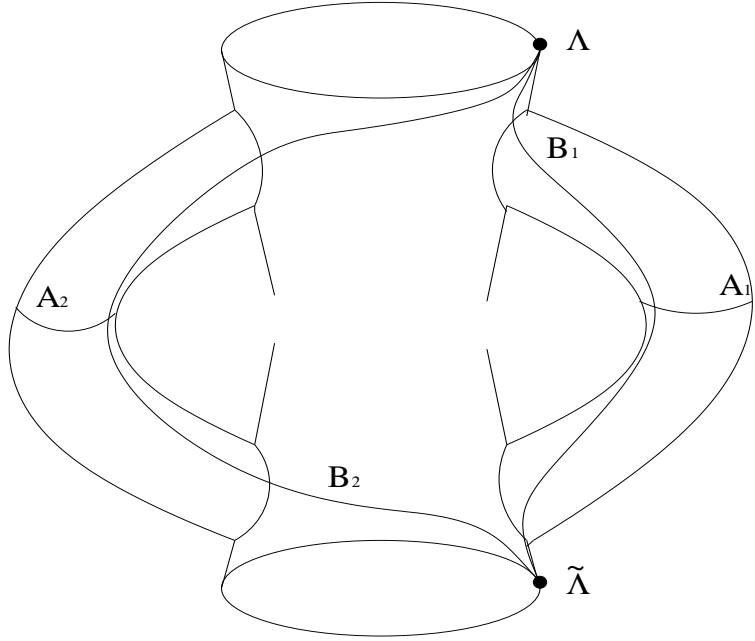,width=12cm,height=14cm}}
\caption{\sl Spectral curve for $S^3/\zet_2$ matrix model.}
\label{plumbing}
\end{figure}

Let's call the contour around the $u$ cut the $A_1$ cycle and the contour around the $\mu$ cut the $A_2$ cycle. There are also two dual $B$ cycles. The $B_1$ cycle starts at a point $\Lambda$ at infinity on the classical sheet where the resolvent is finite and goes to a point $\tilde{\Lambda}$ on the second sheet through the $u$ cut. The end points of the $B_2$ cycle are the same but the contour goes from one sheet to the other through the $\mu$ cut. The Riemann surface is depicted on the figure \ref{plumbing}.  

Now to find $a$, the end point of the $u$ cut, one has to fix the period over the $A_1$ cycle

\be
\frac{1}{4}\oint_{A_1}\omega (z) dz=\pi i S_1.
\ende
Analogously, the period over the $A_2$ cycle must be proportional to $S_2$

\be
\frac{1}{4}\oint_{A_2}\omega (z)dz=\pi i S_2.
\ende
Actually given the normalization condition at $z=\! \! -\infty$, only one of those periods is independent. The integral over the $A=A_1+A_2$ cycle is fixed by

\be
\frac{1}{4} \oint_A\omega (z)dz=-\pi i \tilde{\omega} (-\infty)=\pi i S,
\ende
therefore to fix $a$ we have exactly one integral to do, either the $A_1$ period or $A_2$ period.
These period integrals are  hard to take in an explicit form, we will use a perturbative method 
to calculate them.

The deformed 
CY  is given explicitly from (\ref{w/2}) with $u\equiv z$ and $v=(S-\omega)/2$ as

\be \label{rs_p2}
(e^{v}-1)(e^{2u+v}-1)+e^{S}-1-de^{u+v}=xy
\ende
which is a particular complex structure deformation of $F_{2}=xy$ (from eq. \ref{zpCY}). The mirror of ${\cal O}(-K)\! \! \rightarrow\! \! \projective^{1}\times\projective^{1}$ is given by \cite{Aganagic:2001nx}

\be \label{F_0}
xy=e^{u}+e^{v}+e^{-t-u}+e^{-s-u}+1
\ende
where $t$ and $s$ are complex structure moduli.  There is a simple coordinate transformation that brings (\ref{rs_p2}) to (\ref{F_0}). Explicitly 
$v\rightarrow v+\ln d-S+u+i\pi$ and $u\rightarrow u-\ln d$ which gives us the following relationship  between complex structure moduli and 't Hooft parameters
\be
\label{t}
t=\ln d (S_1,S_2)
\ende
\be
\label{s}
s=2\ln d (S_1,S_2)-S.
\ende

It concludes that the matrix model spectral curve is indeed what we expect from the mirror symmetry. 
In the section 6 we find pertubative expression for the complex structure deformation parameter $d(S_1,S_2)$ and 
for the free energy using the resolvent (\ref{w/2}). Perturbative calculations are valid when the values of 't Hooft parameters are small. Notice that
the coordinate transformation above is not regular when 't Hooft parameters goes to $0$ since $d$ is also small in this limit. Therefore one 
can not use a perturbative expression for $d$ to relate it to the Kahler parameters $Re(t)$ and $Re(s)$ using  (\ref{t}) and (\ref{s}).

%---------------------------------------------------------------------------------------------------------------------------------
\section{General $S^3/\zet_p$ lens spaces.}

We now generalize this analysis to the case $S^{3}/\ZZ_{p}$. Here there are $p$ sets of eigenvalues, we label them by an index $I \in \{0,..,p-1\}$. The measure factor is a product of two factors, a self interacting term $(\Delta_1)$ and a term containing the interaction between different sets of eigenvalues $( \Delta_2)$,

\bea
&&\Delta_1 (u)=\prod_I\prod_{i\ne j}{\left (2\sinh\left (\frac{u^I_i-u^I_j}{2}\right )\right )}^2 \\
&&\Delta_2 (u)=\prod_{I<J}\prod_{i,j}{\left (2\sinh\left (\frac{u_i^I-u_j^J+d^{IJ}}{2}\right )\right )}^2,
\eea
where $d^{IJ}=2\pi i (I-J)/p$. The potential has an overall factor of $p$ compared to the $S^{3}$ case,

\be
V(u)=p\sum_{I,i}\frac{{(u_i^I)}^2}{2}.
\ende
We define individual resolvents for each set of the eigenvalues by

\be
\omega_I (z)=g_s\sum_i\coth\left (\frac{z-u_i^I}{2}\right )
\ende
and the total resolvent, which we are most interested in is 

\be
\omega (z)=\sum_I\omega_I\left (z-\frac{2\pi i I}{p}\right ).
\ende

The equation of motion for each eigenvalue is 

\be \label{zpeom}
p u_{i}^{I}= g_{s}\sum_{i\neq j}\coth\left( \frac{u^{I}_{i}-u^{I}_{j} }{2} \right)
+ g_{s} \sum_{J\neq I} \sum_{j} \coth\left( \frac{u^{I}_{i}-u^{J}_{j}+d^{IJ} }{2} \right).
\ende
From the large $N$ limit of this equation we can derive

\be
\frac{1}{2}\omega^2 (z)-p\sum_I\left (z-\frac{2\pi i I}{p}\right )\omega_I \left (z-\frac{2\pi i I}{p}\right )=f(z),
\ende
where $f(z)$ is a regular function.  From this it follows that

\be
\frac{1}{2}\left( \omega_{+}\left (z+\frac{2\pi i I}{p}\right )+\omega_{-}\left (z+\frac{2\pi i I}{p}\right )\right) =pz,\ \ (I{\rm 'th\ cut}) .
\ende 
and so every cut is indeed a square root. Now we construct a regular function,

\be\label{g_p}
g(Z)=e^{\omega/2}+Z^pe^{-\omega/2},
\ende
which has the limiting behavior,

\bea
&& \lim_{Z\rightarrow \infty}g(Z)=e^{-S/2}Z^{p}\\
&&  \lim_{Z\rightarrow 0}g(Z)=e^{-S/2}
\eea
and is thus of the form,
\be
g(Z)=e^{-S/2}(Z^p+d_{p-1}Z^{p-1}+...+d_{1}Z+1).
\ende
The function $g(Z)$ depends on $p\! -\!1$ moduli $d_{n}$, which could be found by evaluating the period integrals

\be
\frac{1}{2}\oint_{A_I}\omega (z)dz=2\pi iS_I.
\ende
Since we have already fixed the integral over the cycle $A=\sum_I A_I$, there are only $p\! -\! 1$ independent $A$-periods. 

We can solve (\ref{g_p}) for $\omega(Z)$ to get

\be \label{zp_w}
\frac{\omega (Z)}{2}=\log\left (\frac{1}{2}\left (g(Z)-\sqrt{g^2 (Z)-4 Z^p}\right )\right),
\ende
the function under the square root is a polynomial of the degree $2p$, it has $2p$ distinct roots that depend on only $p\! -\! 1$ parameters. Thus the spectral curve consists of two cylinders glued together along $p$ cuts. Note that the center of the $I$'th cut is at the point $z=2\pi i I/p$. From (\ref{zp_w}) we see that the spectral curve is given by

\be
(e^{v}-1)(e^{pu+v}-1)+e^{S}-1+e^{v}\sum_{n=1}^{p-1}d_{n}e^{nu}=0,
\ende
a complex structure deformation of  $F_{p}=0$ (from (\ref{zpCY})).

%---------------------------------------------------------------------------------------------------------------------------------
\section{Free Energy}

An important check of our calculations is to use the resolvent we have found to calculate the free energy perturbatively. In the appendix we perform this for $p=2$, here we quote our result, it agrees with that obtained in \cite{Aganagic:2002wv}. 

From (\ref{gp2}-\ref{xb}) we can see that 

\be \label{w/4}
\frac{\omega(z)}{4}=\log\left (\frac{e^{-S/4}}{2}[\sqrt{(Z+b) (Z+1/b)}-\sqrt{(Z-a)(Z-1/a)}]\right ).
\ende
It is not possible to obtain the parameter $a$ as a explicit function of the 't Hooft parameters but we can find a perturbative series for it. We will do this by introducing two small parameters $\epp$ and $\ep$ in the following way

\be
a+\frac{1}{a}=2 (1+\epp),\hspace{2cm} b+\frac{1}{b}=2 (1+\ep)
\ende
and then performing the $A$ period integrals as an expansion in $\epp$ and $\ep$. This will give the 't Hooft parameters as a power series in $\epp,\ \ep$ which we then invert. We find that

\begin{eqnarray} \label{ep1}
\epp&=&S_1+\frac{1}{4} S_1 (S_1+S_2)+ \\
&+&\frac{1}{96} S_1 (3 S_2^2+9 S_1 S_2+4 S_1^2)+\non \\
&+&\frac{1}{384} S_1 (S_2^3+6S_2^2 S_1+7S_2 S_1^2+2 S_1^3)\non
\end{eqnarray}
and so we see that when $S_{1}=0$, $\epp=0$ and so the second square root in (\ref{w/4}) becomes a complete square thus there is only one cut. This agrees with the fact that if the second cut is empty the problem should reduce to CS theory on $S^{3}$. The corresponding expression for $\ep$ can be obtained from (\ref{epp}) by switching $S_1$ and
$S_2$.

By performing this expansion of the resolvent in $\epp$ and $\ep$ and then calulcating the $B$ period integrals, we can get an expansion for the free energy. This analysis is also done in the appendix, we quote the result

\begin{eqnarray}
\partial_{S_1}F_0(S_1,S_2)&=&-S_1 (1+\log 2)+2S_2\log 2+S_1\log S_1+\frac{1}{8} {(S_1+S_2)}^2+\\
&+&\frac{1}{576} (3 S_2^3+18S_2^2 S_1+9S_2 S_1^2+2S_1^3)+{\cal O}(S^5)
\end{eqnarray}
and there is a similar expression for $\partial_{S_2}F_0$ obtained by switching $S_{1}$ and $S_{2}$. This agrees with the result of \cite{Aganagic:2002wv} where it was calculated using averages in the Gaussian model.

An important check is to see how the above formula reduces to the free energy (\ref{df0s3}) of $S^3$ model if $S_2=0$. This means that the second set of eigenvalues ($\mu_\alpha$'s) disappear and we have the following relationship between the two coupling constants

\be
g_s^{S^3}=\frac{g_s^{S^3/\zet_2}}{2}.
\ende
This leads to 

\be
\partial_S F_0^{S^3}(S)=4\partial_{S_1}F_0^{S^3/\zet_2}(S_1,S_2)|_{S_1=2S,S_2=0}
\ende
which is indeed satisfied. 

%---------------------------------------------------------------------------------------------------------------------------------
\section{Conclusion.}
We have studied the matrix models that describe Chern-Simons theory on the Lens spaces $S^3/\ZZ_p$. We showed that the resolvent has $p$ square root branch cuts and is thus best thought of as a $p$-cut single matrix model. We have found the form of the resolvent and thus the spectral curve. The spectral curve is a $p-1$ genus Riemann surface with 4 points deleted. We would like lend weight to the conjecture that $T^{*}(S^{3}/\ZZ_{p})$ undergoes a large $N$ transition to an $A_{p-1}$ fibration over $\projective^{1}$. We have shown that the spectral curve of the matrix model is topologically equivalent to what is expected from the mirror symmetry. However, even for $p=2$ we have been unable to find an explicit map between Kahler structure moduli of A-model and the periods of the B-model geometry. This must be due to the complexity of the moduli space of the manifold. It would be interesting to understand this better. 

We have also calculated the free energy for the case $p=2$ by keeping one cut small and expanding in the appropriate small parameter. We found agreement with \cite{Aganagic:2002wv} which is a non-trivial check of our resolvent. We also showed that when one cut contains zero eigenvalues that the resolvent and free energy reduce to the case of CS theory on $S^{3}$, which is a further check of our results.

Chern-Simons theory on various manifolds can be described by a matrix model \cite{Marino:2002fk} and the technology introduced in this paper may find applications there. Finally, it it would intriguing if a matrix model description of the topological vertex \cite{Aganagic:2003db} could be found and the matrix models studied in this paper may be a step in that direction.

%---------------------------------------------------------------------------------------------------------------------------------

\begin{acknowledgments}
We would like to thank Mina Aganagic, Kris Kennaway, Christian Romelsberger and Johannes Walcher for useful discussions. Special thanks to Kris Kennaway for a 
help with MATHEMATICA.	
This work was supported in part by funds provided by the DOE under grant number DE-FG03-84ER-40168 and the National Science Foundation under grant number PHY99-07949.
\end{acknowledgments}

%---------------------------------------------------------------------------------------------------------------------------------

\begin{appendix}

\section{Free energy for Chern-Simons on $S^3$.}

In this section we will derive an expression for the leading order free energy ($F_{0}$) in terms of the $B$ period integral for matrix model of CS theory on $S^3$. There is a slight difference here compared to the matrix model with measure on the Lie algebra due to the fact that there the resolvent vanishes at infinity while for our case the resolvent is a non-zero constant at infinity. Nevertheless we still find that $\frac{\del F_{0}}{\del S}$ is proportional to the integral over the B cycle as usual. 

$F_0(S)$ is proportional to the action evaluated on-shell, if we add a single eigenvalue $u$, the free energy changes by

\be
\Delta F_0(S)=-g_s u^2+g^2_s\sum_i \log {\left (2\sinh\left (\frac{u-u_i}{2}\right )\right )}^2
\ende
and the corresponding change in the 't Hooft parameter is $\Delta S=g_s$. Therefore the derivative of
the free energy with respect to $S$ is

\be \label{deltaFS3}
\partial_S F_0(S)=\frac{\Delta F_0}{g_s}.
\ende

Let's take a point at infinity $\Lambda$, then the following relation holds

\be\label{rel}
\log {\left (2\sinh\left (\frac{u-u_i}{2}\right )\right )}^2=-{\rm P}\int^\Lambda_u\coth\frac{z-u_i}{2}dz-u_i,
\ende
where all terms except finite ones have been dropped. The last term is not present for Lie algebra matrix models, here it is due to the fact that the resolvent is finite at $\Lambda$. However, because (\ref{EM}) summed over $i$ gives 

\be \label{sum}
\sum_i u_i=0
\ende
the last term in (\ref{rel}) vanishes when summed over $i$. Now it is easy to recognize the integral over the resolvent in (\ref{deltaFS3}) 

\be
\partial_S F_0(S)=\int^\Lambda_u dz (z-\omega (z))=-\oint_B\frac{\omega (z)}{2} dz.
\ende
The integral can be taken explicitly with the result 

\be
\label{df0s3}
\partial_S F_0(S)=-\frac{\pi^2}{6}+\frac{S^2}{2}+{\rm Li}_2 \left (e^{-S}\right ),
\ende
where ${\rm Li}_2 (x)$ is the Euler's dilogarithm function 

\be
{\rm Li}_2(x)=\sum^\infty_{n=1}\frac{x^n}{n^2}.
\ende
We will need this result when we calculate the corresponding free energy for CS theory on $S^3/\ZZ_2$.  

%-------------------------------------------------------------------------------------------

\section{Free energy for $S^3/\zet_2$.}

We now extend this analysis to Lens spaces. Let's look at how the action changes if one eigenvalue is added. Then we divide that change by $g_s$, which is the corresponding change in the 't Hooft parameter $S_1$, use (\ref{deltaFS3}) and the identity

\be
\log {\left (2\cosh\left (\frac{u-\mu_\alpha}{2}\right )\right )}^2=-{\rm P}\int^\Lambda_u\tanh\frac{z-\mu_\alpha}{2}dz-\mu_\alpha,
\ende
\be
\sum_i u_i+\sum_\alpha\mu_\alpha =0.
\ende
In this way the following expression is obtained

\be
\label{dfz2}
\partial_{S_1}F_0(S_1,S_2)=-2\oint_{B_1}\frac{\omega (z)}{4}dz\equiv -2\Pi_1
\ende

Analogously, in order to get the derivative of the free energy with respect to $S_2$ the integral over the $B_2$-cycle has to be taken. We found the resolvent to be 

\be
\label{omegaba} 
\frac{\omega(z)}{4}=\log\left (\frac{M}{2}[\sqrt{(Z+b) (Z+1/b)}-\sqrt{(Z-a)(Z-1/a)}]\right )
\ende
with $M=e^{-S/4}$ and 

\be
\label{ba}
(Z+b) (Z+1/b)=(Z-a)(Z-1/a)+4ZM^{-2}.
\ende

There is one more parameter to fix, the end point $a$ of the $u$ cut. The contour 
integral over one of the $A$-cycles must be equal to the corresponding 't Hooft parameter. We consider the $A_1$-cycle

\be
\label{A1}
\oint_{A_1}\frac{\omega (z)}{4}dz=\pi i S_1.
\ende

The cycle $A=A_1+A_2$ can be deformed to a contour around the logarithmic cut in the $Z$ variable, which is proportional to the value of the other branch of the resolvent at zero, $\tilde{\omega} (0)=-S$. Note that the resolvent on the second sheet is the same as in (\ref{omegaba}) except that there is a plus sign between the two square roots. So the period integral over the $A$-cycle is $\pi i S$. Alternatively, one can take periods over $A_1$ and $A_2$ cycles as independent ones and obtain that $\tilde{\omega} (0)=-S$ as a consequence. The problem of calculating integrals like (\ref{A1}) is very similar to the case of Lie algebra matrix models, in both cases they can be reduced to elliptic integrals. It is not possible to obtain the end point of the cut $a$ as an explicit function of the 't Hooft parameters $S_1$ and $S_2$, however if the size of the cuts is small then one can expand in a power series of this small parameter much like the solution of two cut Lie algebra matrix models. To this end it is better to make $a$ and $b$ being independent, fix the $A_1$ and $A_2$ periods and recover a perturbative analog of the relation (\ref{ba}). 

So we introduce two small parameters $\epp$ and $\ep$ in the following way

\be
a+\frac{1}{a}=2 (1+\epp),\hspace{2cm} b+\frac{1}{b}=2 (1+\ep).
\ende
The resolvent up to a nonsingular term becomes

\be
\frac{\omega(z)}{4}\sim\log\left (\sqrt{Z^2+2Z(1+\ep)+1}-\sqrt{Z^2-2Z(1+\epp)+1}\right ).
\ende

To take the integral (\ref{A1}) one first expands $\omega/4$ around $\ep=0$ keeping the size of the $A_1$ cycle finite. One square root disappears so the integrals become tractable. The method is similar to one used in 
\cite{Cachazo:2001jy}.
Note that at $\ep=0$ one cut shrinks to the zero size and  the resolvent matches one of Chern-Simons on $S^3$ matrix models. We expand up to the fourth power in $\ep$ and $\epp$,

\bea \label{S1ser}
S_1&=&4\log\mu (\epp)+\frac{\ep}{2\mu(\epp)}\A_1-\\
&-&\frac{\ep^2}{8}\left (\frac{1}{2\mu^2(\epp)}\A_1+\frac{1}{\mu(\epp)}\A_2\right )+\nonumber\\
&+&\frac{\ep^3}{24}\left (\frac{1}{2\mu^3(\epp)}\A_1+\frac{1}{\mu^2(\epp)}\A_2+\frac{3}{\mu(\epp)}\A_3\right )-\nonumber\\
&-&\frac{\ep^4}{32}\left (\frac{5}{\mu(\epp)}\A_4+\frac{3}{2\mu^2(\epp)}\A_3+\frac{1}{2\mu^3(\epp)}\A_2 + \frac{3}{8\mu^4(\epp)}\A_1\right )\nonumber
\eea
with $\mu(\epp)=1+\epp/2$ and

\bea
&& \A_1=\frac{1}{\pi i}\oint_{A_1}\frac{\sqrt{(Z-a)(Z-1/a)}}{1+Z}\frac{dZ}{Z}=4 (1-\sqrt{\mu(\epp)}),\\
&& \A_2=\frac{1}{\pi i}\oint_{A_1}\frac{\sqrt{(Z-a)(Z-1/a)}}{(1+Z)^3}dZ=-\frac{\epp}{4\sqrt{\mu(\epp)}}, \\
&& \A_3=\frac{1}{\pi i}\oint_{A_1}\frac{Z\sqrt{(Z-a)(Z-1/a))}}{(1+Z)^5}dZ=-\frac{\epp (8+3\epp)}{128\mu^{3/2}(\epp)}, \\
&& \A_4=\frac{1}{\pi i}\oint_{A_1}\frac{Z^2\sqrt{(Z-a)(Z-1/a)}}{(1+Z)^7}dZ.
\eea 

Since $A_4$ shows up in the fourth order we only need its value at $\epp=0$ which is zero. The term of order ${\cal O}(\ep^0)$ has been calculated using the known expression of $S^3$ case. The next step is to expand in $\epp$

\begin{eqnarray} \label{S1}
S_1&=&\epp-\frac{1}{4}\epp (\ep+\epp)+ \\
&+&\frac{1}{96}\epp (9\ep^2+15\epp\ep+8\epp^2)-\non \\
&-&\frac{1}{128}\epp (5\ep^3+12\ep^2\epp+11\ep\epp^2+4\epp^3).\non 
\end{eqnarray}

To find a similar expression for the $A_2$ period one just has to replace $S_1$ by $S_2$ and switch $\epp$ and $\ep$ in the above formula. An important check is to recover relation (\ref{ba}) written in the form

\be
S=S_1+S_2=2\log\left (1+\frac{\ep+\epp}{2}\right )
\ende
and expanded up to the fourth power in $\ep + \epp$. 

The two power series for the two 't Hooft parameters can be inverted giving 

\begin{eqnarray} \label{epp}
\epp&=&S_1+\frac{1}{4} S_1 (S_1+S_2)+ \\
&+&\frac{1}{96} S_1 (3 S_2^2+9 S_1 S_2+4 S_1^2)+\non \\
&+&\frac{1}{384} S_1 (S_2^3+6S_2^2 S_1+7S_2 S_1^2+2 S_1^3).\non
\end{eqnarray}
The corresponding series for $\ep$ can be obtain from the above expression by switching $S_1$ and $S_2$. 

In a similar fashion one can calculate periods over the $B$ cycles. Let's find the period $\Pi_1$ over the $B_1$ cycle

\be
\Pi_1=\int^{\Lambda}_{\widetilde{\Lambda}}\frac{\omega (Z)}{4}\frac{dZ}{Z},
\ende
where $\Lambda$ is a point at infinity on the first sheet and $\widetilde{\Lambda}$ is a point at infinity on the second sheet. Again, the first step is to expand the resolvent in power series of $\ep$. The integral in the ${\cal O}(\ep^0)$ term has been taken using the known result from CS theory on $S^3$. 

\begin{eqnarray}
\Pi_1&=&\frac{\pi^2}{6}-\frac{1}{2}\log^2\mu(\epp)-{\rm Li}_2 \left (\frac{1}{\mu(\epp)}\right )+\frac{\ep}{2\mu(\epp)}\B_1-\\
&-&\frac{\ep^2}{8}\left (\frac{1}{2\mu^2(\epp)}\B_1+\frac{1}{\mu(\epp)}\B_2\right )+\nonumber\\
&+&\frac{\ep^3}{24}\left (\frac{1}{2\mu^3(\epp)}\B_1+\frac{1}{\mu^2(\epp)}\B_2+\frac{3}{\mu(\epp)}\B_3\right )-\nonumber\\
&-&\frac{\ep^4}{32}\left (\frac{5}{\mu(\epp)}\B_4+\frac{3}{2\mu^2(\epp)}B_3+\frac{1}{2\mu^3(\epp)}\B_2+
\frac{3}{8\mu^4(\epp)}\B_1\right ),\nonumber
\end{eqnarray}
where

\bea
&& \B_1=\int_{\widetilde{\Lambda}}^\Lambda\frac{\sqrt{(Z-a)(Z-1/a)}}{1+Z}\frac{dZ}{Z}, \\
&& \B_2=\int_{\widetilde{\Lambda}}^\Lambda\frac{\sqrt{(Z-a)(Z-1/a)}}{(1+Z)^3}dZ, \\
&& \B_3=\int_{\widetilde{\Lambda}}^\Lambda \frac{Z\sqrt{(Z-a)(Z-1/a)}}{(1+Z)^5}dZ, \\
&& \B_4=\int_{\widetilde{\Lambda}}^\Lambda \frac{Z^2\sqrt{(Z-a)(Z-1/a)}}{(1+Z)^7}dZ.
\eea
Then one sends $\Lambda$ to infinity and takes the finite part, which is then expanded in powers of $\epp$

\begin{eqnarray}
\B_1&=&-\log 16+\frac{1}{2} (-1+\log (\epp/8))\epp +\frac{1}{32} (-1-2\log (\epp/8))\epp^2+\\
&+&\frac{1}{384} (5 +6\log(\epp/8))\epp^3+{\cal O}(\epp^4),\non
\end{eqnarray}
\bea
&& \B_2=\frac{1}{2}+\frac{1}{8}\epp\log (\epp/8)-\frac{1}{32} (1+\log (\epp/8))\epp^2+{\cal O}(\epp^3), \\
&& \B_3=\frac{1}{16}+\frac{1}{64} (1-2\log(\epp/8))\epp +{\cal O}(\epp^2), \\
&& \B_4=\frac{1}{96}+{\cal O} (\epp).
\eea
Combining this all together and plugging in the expressions for $\epp$ and $\ep$ as functions of $S_1$ and $S_2$ and using (\ref{dfz2}) we get

\begin{eqnarray} \label{pi}
\partial_{S_1}F_0(S_1,S_2)&=&-S_1 (1+\log 2)+2S_2\log 2+S_1\log S_1+\frac{1}{8} {(S_1+S_2)}^2+\\
&+&\frac{1}{576} (3 S_2^3+18S_2^2 S_1+9S_2 S_1^2+2S_1^3)+{\cal O}(S^5),
\end{eqnarray}
which is in agreement with \cite{Aganagic:2002wv}. Note there is no terms of order ${\cal O} (S^4)$.

\end{appendix}

%-------------------------------------------------------------------------------------------------------------------------------------

\end{document}